\documentclass[12pt]{article}
\usepackage{amsmath}
\usepackage{citesort}
\usepackage{amssymb}
\usepackage{epsfig}
\def \BE {\begin{equation}}
\def \EE {\end{equation}}
\def \BEA {\begin{eqnarray}}
\def \EEA {\end{eqnarray}}

\begin{document}

\title{Differential approximation for Kelvin-wave turbulence.}
\author{Sergey Nazarenko
\\ \ \\
{Mathematics Institute,  University of Warwick, Coventry CV4 7AL, UK}}
\maketitle

\begin{abstract}
I present a nonlinear differential equation model (DAM) for
the spectrum of Kelvin waves on a thin vortex filament.
This model preserves the original scaling of the
six-wave kinetic equation, its direct and inverse cascade
solutions, as well as the thermodynamic equilibrium spectra.
Further, I extend DAM to include the effect of sound radiation
by Kelvin waves. I show that, because of the phonon
radiation, the turbulence spectrum ends at a maximum frequency
$\omega^* \sim  (\epsilon^3 c_s^{20} / \kappa^{16})^{1/13}$ where $\epsilon$ is the total
energy injection rate, $c_s$ is the speed of sound and $\kappa$ is the
quantum of circulation.

\end{abstract}

\section{Kelvulence: cascades and spectra.}

Kelvin waves propagating on a thin vortex filament were proposed by Svistunov
to be a vehicle for the turbulent cascades in superfluids
near zero temperature  \cite{svist}. 
Presently it is a widely accepted view well supported by the theory
and numerical simulations, see e.g. \cite{vinen,kivot,ks,vtm,ks1}.
I will refer to the state characterised by random nonlinearly interacting
Kelvin waves as ``kelvulence'' (i.e. Kelvin turbulence).
Recently, Kozik and Svistunov \cite{ks} used the weak turbulence
approach to kelvulence and derived a six-wave kinetic equation (KE) for the
spectrum of weakly nonlinear Kelvin waves. Based on KE, they derived a
spectrum of waveaction that corresponds to the constant Kolmogorov-like
cascade of energy from small to large wavenumbers,
\begin{equation}
n_k \sim k^{-17/5}.
\label{direct}
\end{equation}
Because the number of waves in the leading resonant process is even (i.e. 6),
KE conserves not only the total energy but also the total waveaction of the system.
The systems with two positive conserved quantities are known in turbulence 
to possess a dual cascade behavior. For the Kelvin waves, besides the direct
energy cascade there also exists an inverse cascade of waveaction, the spectrum
for which was recently found by Lebedev \cite{leb},
\begin{equation}
n_k \sim k^{-3}.
\label{inverse}
\end{equation}
Interestingly, such $-3$ spectrum was suggested before by Vinen based on a 
dimensional argument not involving the energy flux  \cite{vinen}.
Similar argument in water wave turbulence gives famous Phillips spectrum which 
is associated with sharp water crests due to wavebreaking which occurs at large excitation
levels. By analogy, we could expect that Vinen $-3$ spectrum should be observed 
in kelvulence at high excitation levels leading to sharp angles due to reconnections. 
This view is supported by recently reported  numerics by Vinen, Tsubota and Mitani \cite{vtm},
where it was argued that the $-3$ exponent arises when the vortex line bending angle becomes
large (of order one).
Kozik and Svistunov \cite{ks1} later reported a  result obtained with a refined numerical method which gave
a spectrum closer to the $-17/5$ shape, which is a Kolmogorov-like direct cascade of energy
dominating the wavebreaking effects at lower turbulence levels.
The inverse cascade spectrum, although it has the same $-3$ exponent as the Vinen spectrum,
is fundamentally different: it corresponds to weak rather than strong turbulence and is
more relevant if the main source of the Kelvin waves is at small scales.
For example,  the smallest scales on the vortex filament could be generated by 
reconnections via sharp angles produced by these processes.

\section{Differential equation model}

Differential equation models proved to be very useful for analysis in both
weak turbulence \cite{irosh,hass,sand} and strong turbulence \cite{leith,cn,lnv}.
These equations are constructed in such a way that they preserve the main scalings
of the original closure (KE in the case of weak turbulence), in particular,
its nonlinearity degree with respect to the spectrum and its cascade and thermodynamic solutions.
For the Kelvin wave spectrum these requirements yield,
\begin{equation}
\dot n = {C  \over \kappa^{10}} \omega^{1/2} {\partial^2 \over \partial \omega^2}
\left(
 n^6 \omega^{21/2} {\partial^2 \over \partial \omega^2} {1 \over n}
\right),
\label{forth}
\end{equation}
where $\kappa$ is the vortex line circulation, $C$ is a dimensionless constant
and $\omega = \omega (k) = {\kappa \over 4 \pi} k^2$ is the Kelvin wave
frequency (here, we ignore logarithmic factors).
This equation preserves the energy
\begin{equation}
E = \int \omega^{1/2} n \, d\omega
\end{equation}
and the waveaction
\begin{equation}
N = \int \omega^{-1/2} n \, d\omega.
\end{equation}
It is an easy calculation to check that equation (\ref{forth}) has both the
direct cascade solution (\ref{direct}) and  the inverse cascade solution
 (\ref{inverse}).
It also has the same thermodynamic Rayleigh-Jeans solutions as the original KE,
\begin{equation}
n = { T \over  \omega + \mu}.
\label{term}
\end{equation}
where $T$ and $\mu$ are constants having a meaning of temperature and the chemical
potential respectively. 

\section{Kelvulence radiating sound.}

In contrast with the classical Navier-Stokes flow, there is no viscosity that
could dissipate the superfluid turbulent cascade at small scales.
At low temperatures, friction with the normal component is also inefficient
and the only dissipative process which can absorb the cascade is radiation
of sound by moving superfluid vortex filaments \cite{vinen}.
Let introduce the sound dissipation effect into DAM by using the classical
results of Lighthill about sound produced by classical turbulence \cite{lighthill}.
Namely, we will use the result that the rate at which the sound energy is
generated is proportional to the fourth power of the Mach number $M = v/c_s \sim n^{1/2}/c_s$,
where $v$ is characteristic velocity in turbulence and $c_s$ is the speed of
sound.  Further,
we will ignore the logarithmic corrections due to the finite vortex core size $a$.
In the other words, $a$ should not enter the expression explicitly
but only implicitly via $c_s \sim \kappa/a$ (which in turn enters only via $M$).
Then the rest of the sound dissipation term can be completed uniquely via 
the dimensional argument and the result is
\begin{equation}
(\dot n)_{radiation} = - { \omega^{9/2} n^2 \over  \kappa^{1/2} c_s^4}.
\label{soundrad}
\end{equation}
Let us now examine what effect of the sound radiation on the direct energy cascade.
For these purposes we can consider even simpler first order DAM which still describes
the direct cascade but ignores the inverse cascade and the thermodynamic solutions.
Such DAM, including the sound radiation term, reads:
\begin{equation}
\dot n = {C_1  \over \kappa^{10}} \omega^{-1/2} {\partial \over \partial \omega}
\left(
 n^5 \omega^{17/2}
\right)
- { C_2 \omega^{9/2} n^2 \over  \kappa^{1/2} c_s^4},
\label{first}
\end{equation}
where $C_1$ and $C_2$ are dimensionless constants.
The general stationary solution of equation (\ref{first}) is
\begin{equation}
 n = A \left( 1- B \omega^{13/5} \right)^{1/3} \omega^{-17/10},
\label{radspec}
\end{equation}
where 
\begin{eqnarray}
A &=& (2 /C_1)^{1/5} \kappa^{21/10} \epsilon^{1/5}, \\
B &=& {3 \cdot 2^{-3/5} \over 13}  C_2 \kappa^{16/5} c_s^{-4} \epsilon^{3/5},  
\end{eqnarray}
and $\epsilon$ is the total energy dissipated in the system per unit time per unit
length.

In the absence of sound radiation, $B=0$, we recover the direct cascade spectrum
(\ref{direct}). For  $B > 0$, the spectrum has the direct cascade shape
(\ref{direct}) at low frequencies, $\omega \ll B^{-5/13}$, and it falls to zero
at a finite frequency 
\begin{equation}
\omega^* \sim  B^{-5/13} \sim (\epsilon^3 c_s^{20} / \kappa^{16})^{1/13}.   
\label{cutoff}
\end{equation}
On the other hand, Kelvin waves can only have wavelengths greater than the
vortex radius $a$ which in terms of the frequency means $\omega < \omega_c = c_s^2/\kappa$.
Thus, expression (\ref{radspec}), particularly the finite cut-off at $\omega^*$, will only
hold if $\omega^* < \omega_c$ or
\begin{equation}
\epsilon < \kappa c_s^{2}.
\end{equation}
This condition can be formulated in terms of the characteristic bending angle of the vortex
line $\alpha$ which in the direct cascade state is related to $\epsilon$ as
\begin{equation}
\alpha \sim (\epsilon l^2 /\kappa^3)^{1/10}, 
\end{equation}
where $l$ is characteristic length of the Kelvin waves.
Thus, condition $\omega^* < \omega_c$ becomes
\begin{equation}
\alpha <  (l/a)^{1/5},
\end{equation}
which always holds for $\alpha \lesssim 1 $ because $l>a$.
Therefore we conclude that the kelvulence cascade will always
decay to zero before reaching the maximal allowed frequency 
of the propagating Kelvin waves.

\section{Conclusions.}

In this Letter I presented a differential equation model (DAM) for 
the Kelvin wave turbulence (kelvulence) consistent with the
scalings of, and having the same
set of cascade and thermodynamic solutions
as, the original six-wave kinetic equation.
DAM also exists in reduced versions if some of the flux and thermal
solutions are not important for a particular problem and can be ignored
(in this Letter I presented three versions given by the fourth, the second and
the first order equations respectively).

Based on the Lighthill's theory of sound produced by turbulence,
I extended DAM to include the effect of kelvulence dissipation by
 sound radiation. I obtained a steady state solution of this model corresponding
to the direct cascade of wave energy gradually dissipated via radiation
and decaying to zero at a finite wavenumber $\omega^*$ given by the
expression (\ref{cutoff}).
This solution can also be used to predict the spectrum of the radiated
sound, because the sound energy is just the same as the energy lost
by kelvulence via radiation.
One can also extend this study to scattering of sound by turbulence
in systems where such sound is generated by externally.

Finally, in appendix I presented a ``warm cascade'' solution for kelvulence which
is relevant for numerical simulations with a finite cut-off frequency.

I thank Carlo Barenghi, Vladimir Lebedev, Boris Svistunov, Joe Vinen and Grisha Volovik
for their helpful comments about my model.

\section{Appendix: Warm cascade solutions.}

Above, we ignored the ``thermal'' component in turbulence
which is totally justified because this component does not show
up when turbulence is dissipated gradually in the Fourier space,
as it is done via the sound radiation in our case. On the other hand,
it is known in the turbulence theory that sharp dissipation or presence of a 
cut-off frequency lead to reflection
of a large portion of the energy flux from the smallest scale and, consequently, an
pile-up of the spectrum near the smallest scale. This ``bottleneck'' effect can be described
in terms of ``warm cascade'' solutions in which both an energy flux and a thermal
components are present.  As I showed above, the natural cut-off frequency of Kelvin
waves $\omega_c$ is not going to be reached by the turbulent cascade because it will
always terminate at a frequency $\omega_* < \omega_c$ due to the sound radiation.
However, in numerical simulations of kelvulence the cut-off frequency may be 
less than $\omega_*$ due to the limited numerical resolution. Thus, the warm cascade
solutions could be relevant for understanding spectra obtained numerically.
Such warm cascade solutions were originally obtained for
classical Navier-Stokes turbulence in \cite{cn}
and here we will follow a similar approach to find such states for kelvulence.
To describe both the energy cascade and the thermodynamic component (but still ignore
the inverse cascade solution) we need to use the second order version of DAM, namely
\begin{equation}
\dot n = {C  \over \kappa^{10}} \omega^{-1/2} {\partial \over \partial \omega}
\left(
 n^4 \omega^{17/2} {\partial (\omega n) \over \partial \omega}
\right).
\label{second}
\end{equation}
The general steady state solution of this equation is
\begin{equation}
n = \omega^{-1} \left( {20 \over 7 C} \kappa^{21/2} \epsilon \omega^{-7/2} + T^5 \right)^{1/5},
\label{warm}
\end{equation}
where $T$ is an arbitrary constant having the meaning of ``temperature''.
This is the  ``warm cascade'' solution
which has the pure cascade solution and the thermodynamic energy equipartition
as two of its limits.
Note that the thermodynamic part of this solution describes
the spectrum pile-up due to the bottleneck phenomenon in numerical
simulations due to presence of a cutoff frequency.
The relative strength of the cascade and the thermodynamic components
will be determined by the ratio of the incident and the reflected energy
fluxes which, in turn, will depend on the particular form of dissipation
at small scales chosen in numerical simulations.
The goal of the true numerical simulation is to chose the dissipation function
in such a way that the bottleneck effect is minimal or absent.

\end{document}